\documentclass[11pt]{article}

\usepackage{amsthm,amsmath}

\usepackage[utf8]{inputenc} %unicode support
\usepackage[numbers]{natbib}
\usepackage{epstopdf}
\usepackage{amsbsy}
\usepackage{amssymb}
\usepackage{amsmath}

\usepackage{geometry}
\def\includegraphics{}
\newcommand{\Xg}{\boldsymbol{X}^T_i\gamma}
\newcommand{\eXg}{\exp (\boldsymbol{X}^T_i\gamma)}
\newcommand{\X}{\boldsymbol{X}^T_i}
\newcommand{\denomS}{(\eXg-1) \, b_i +1}
\newcommand{\denomL}{D_1^2 \,-4\,\rho\eXg \, b_i \,(1-b_i)}
\newcommand{\XtX}{\boldsymbol{X}_i \X}

\begin{document}

\title{A Statistical Model for the Analysis of Beta Values in DNA Methylation Studies}

\author{Leonie Weinhold$^1$, Simone Wahl$^2$, Matthias Schmid$^1$\\[.15cm] \footnotesize
$^1$Department of Medical Biometry, Informatics and Epidemiology\\[-.15cm] \footnotesize
University of Bonn,
Sigmund-Freud-Str.\@ 25,
D-53127 Bonn,
Germany\\[-.15cm] \footnotesize
Email: weinhold@imbie.uni-bonn.de\\ \footnotesize
$^2$Research Unit of Molecular Epidemiology, 
Helmholtz Zentrum M\"unchen\\[-.15cm] \footnotesize
Ingolst\"adter Landstr.\@ 1,
D-85764 Neuherberg, Germany
}

\date{}

\maketitle

\begin{abstract} % abstract
\noindent {\em Background:}
The analysis of DNA methylation is a key component in the development of personalized treatment approaches. A common way to measure DNA methylation is the calculation of beta values, which are bounded variables of the form~$M/(M+U)$ that are generated by Illumina's 450k BeadChip array. The statistical analysis of beta values is considered to be challenging, as traditional methods for the analysis of bounded variables, such as M-value regression and beta regression, are based on regularity assumptions that are often too strong to adequately describe the distribution of beta values. \newline
{\em Results:} %if any
 We develop a statistical model for the analysis of beta values that is derived from a bivariate gamma distribution for the signal intensities $M$ and $U$. By allowing for possible correlations between $M$ and $U$, the proposed model explicitly takes into account the data-generating process underlying the calculation of beta values. \newline
{\em Conclusion:} %if any
The proposed model can be used to improve the identification of associations between beta values and covariates such as clinical variables and lifestyle factors in epigenome-wide association studies. It is as easy to apply to a sample of beta values as beta regression and M-value regression.
 
\end{abstract}

\section{Background}
\label{s1}

The analysis of DNA methylation has become of considerable interest in biomedical research, as epigenetic studies have shown numerous associations between methylation levels and diseases such as cancer  and cardiovascular disease \cite{bird, shames, Sarkar, varley, cancerPurity}. Today, most research focuses on the cytosine-guanine dinucleotide (``CpG'') sites of the DNA, which are the locations where methylation is primarily found in humans \cite{portela}. One of the most widely used techniques to measure DNA methylation is the Illumina Infinium HumanMethylation450 BeadChip array, which covers approximately 450,000 CpG sites. At each CpG site, methylation is quantified by the beta value $b := M / (M +  U + a)$, where $M > 0$ and $U > 0$ denote the methylated and unmethylated signal intensities, respectively, measured by the Illumina 450k array. The offset $a \ge 0$ is usually set equal to 100 and is added to $M + U$ to stabilize beta values when both $M$ and~$U$ are small.

An important goal of methylation analysis is to identify DNA regions where methylation is associated with disease status, lifestyle factors and other clinical or sociodemographic variables \cite{dedeur, Wahletal, dolz, singman}. This is often achieved by fitting site-wise regression models with dependent variable $b$ and a vector of covariates $\boldsymbol{X}$ that may also include potential confounders. After model fitting, a common strategy is to carry out downstream hypothesis tests to identify those CpG sites that show significant associations between methylation status and the variables of interest.

Because, by definition, $b$ is bounded between 0 and 1, Gaussian regression with untransformed beta values is problematic in the context of DNA methylation analysis. In particular, the variance of~$b$ is usually smaller near the boundaries than near the middle of the interval (0,1), implying that the homoscedasticity assumption in Gaussian regression is violated \cite{Duetal, Laird, boostedBeta}. To address this problem, several modeling strategies have been developed, including Gaussian regression with logit-transformed beta values (``M-values'', \cite{Duetal}) and generalized regression models for untransformed bounded responses, e.g.\@ beta regression \cite{ferrari}. Regarding the analysis of DNA methylation, both strategies are intrinsically problematic: In case of M-value regression, the assumptions of a Gaussian model are often not met despite the transformation of the data, and the interpretation of the coefficient estimates is only possible on the transformed scale but not on the original scale of $b$ \cite{ferrari, bock}. Beta regression, on the other hand, requires the ratio $M / (M + U + a)$ to follow a beta distribution, implicitly assuming that the variables $M$ and $U$ are independently gamma distributed \cite{devroye}. While $M$ and $U$ can indeed be described by gamma distributed random variables \cite{Laird2, benner}, the independence assumption for the two signal intensities is often not met in practice. For example, Laird~\cite{Laird} reported that the methylated and unmethylated signal intensities, as produced by the Illumina 450k array, are usually positively correlated. These issues, along with the results of two recent empirical studies \cite{benner, Wahletal}, suggest that both M-value regression and beta regression need to be improved to describe the distribution of $b$ in a statistically sound way.

To address this problem, we propose a novel analysis technique for beta values that relaxes the independence assumption between the signal intensities $M$ and $U$. The idea is to start with a model for the bivariate distribution of $M$ and $U$ and to derive the probability density function of the ratio $M / (M + U)$. This  function is subsequently used to construct the log-likelihood function of a generalized regression model that relates beta values to linear functions of the covariates. Because estimation of the model parameters is based on the maximum likelihood principle, asymptotic confidence intervals and normally distributed test statistics can be derived by evaluating the inverse of the observed information matrix. This strategy allows for downstream hypothesis tests on the associations between a covariate of interest and the methylation status at individual CpG sites. For the rest of this paper, we will refer to the proposed model as ``RCG'' ({\em R}atio of {\em C}orrelated {\em G}ammas)~model.

	\section{Methods and Results}
	\label{s2}
	
	In Section \ref{s21} we introduce basic notation and definitions. Section \ref{s22} briefly reviews beta regression and M-value regression and discusses the limitations of the two methods. In Section \ref{s23} the proposed RCG model for the analysis of beta values is derived. Section \ref{s24} provides details on model fitting and on the construction of downstream hypothesis tests.
	
	\subsection{Notation and Definitions}
	\label{s21}
	
	At each CpG site, the Illumina 450k array produces a sample of methylated and unmethylated signal intensities $(M_i,U_i)_{i=1,\ldots , n}$, where $n$ is the number of analyzed persons. The corresponding set of beta values is calculated by $b_i = M_i / (M_i +U_i+ a)$, $i=1,\ldots , n$. To facilitate the derivation of distributional results, we will set $a = 0$ throughout this section. The predictor variable(s) of interest and the confounding variables are collected in vectors $\boldsymbol{X}_i = (1, X_{i1}, \ldots , X_{ip})^\top$, $i=1,\ldots , n$. For each CpG site, the aim is to analyze the associations between the variables in $\boldsymbol{X}$ and the methylation status~$b$.
	
	Following \cite{Laird2} and \cite{benner}, we assume that the stochastic behavior of the signal intensities $M$ and $U$  can be described by gamma distributed random variables with densities
	\begin{eqnarray}
	\label{gammas}
	f_M (m) &=& \frac{\lambda_m}{\Gamma (\alpha_m )} \left( \lambda_m m \right)^{\alpha_m - 1} \exp (-\lambda_m m) \, , \\ 
	f_U (u) &=& \frac{\lambda_u}{\Gamma (\alpha_u )} \left( \lambda_u u \right)^{\alpha_u - 1} \exp (-\lambda_u u) \, , 
	\end{eqnarray}
	where $\alpha_m, \alpha_u$ and $\lambda_m, \lambda_u$ are the shape and rate parameters of $f_M$ and~$f_U$, respectively. From~(\ref{gammas}) it follows that the means and variances of $M$, $U$ are given by $\alpha_m / \lambda_m$, $\alpha_u / \lambda_u$ and $\alpha_m / \lambda_m^2$, $\alpha_u / \lambda_u^2$, respectively \cite{contbivar}.

	\subsection{Regression Models for the Analysis of Beta Values}
	\label{s22}
	
	Since the ratio $b = M / (M + U) $ is bounded between 0 and 1, it has been argued that a linear regression model of the form
	\begin{equation}
	b = \boldsymbol{X}^\top \gamma + \epsilon   \, ,  \ \ \,  \gamma \in \mathbb{R}^{p+1}   \, ,  \ \ \,  \epsilon \sim \mathcal{N}(0, \sigma^2) \, ,
	\end{equation}
	is not appropriate to model DNA methylation. In particular, the variance of $b$ is usually smaller near the boundaries than near the middle of the interval (0,1), implying that the homoscedasticity assumption $\text{var} (\epsilon) = \sigma^2$ is violated \cite{Duetal}. 
	
	In view of this problem, several statistical models for bounded response variables have been developed (see \cite{boostedBeta} for an overview). A simple approach is to calculate logit-transformed beta values (``M-values'', \cite{Duetal}) and to fit a linear regression model of the form
	\begin{equation}
	\log_2 \left( \frac{b}{1 - b} \right) =  \boldsymbol{X}^\top \gamma + \epsilon \, , \ \ \, \epsilon \sim \mathcal{N}(0, \sigma^2) \, .
	\end{equation}
	Although this strategy has become popular in the analysis of DNA methylation, it has the drawback that the methylation status (as quantified by the value of $b$) is not analyzed on its original scale but on a transformed scale \cite{ferrari}. Furthermore, as shown by Wahl {\it et al.} \cite{Wahletal}, the empirical distribution of logit-transformed beta values usually deviates from normality.
	
	An alternative approach that operates on the untransformed scale of $b$ is {\em beta regression}, which is characterized by a beta distributed outcome variable with probability density function
	\begin{equation}
	\label{betadensity}
	\varphi (b ) =\, \frac{\Gamma (\phi) }{\Gamma (\mu \phi) \Gamma ((1-\mu)\phi )}\,
	b^{\mu\phi - 1}\, (1-b )^{(1-\mu )\phi - 1} \, ,
	\end{equation}
	where $\mu$ and $\phi $ denote the mean and precision parameters, respectively, of the probability density function $\varphi$.
	The predictor-response relationship is usually defined by a monotone increasing link function $g(\cdot )$ and by the model equation $g( \mu |\boldsymbol{X}) = \boldsymbol{X}^\top \gamma$ \cite{ferrari}. A common choice for $g$ is the logit transformation $\log (\mu / (1-\mu))$.
	Since the variance of a beta distributed random variable is given by $\mu (1-\mu) / (1+\phi)$, beta regression accounts for heteroscedasticity and for small variances near the boundaries of the interval (0,1). On the other hand, a major shortcoming of (\ref{betadensity}) in the context of DNA methylation analysis is that
	the signal intensities $M$ and $U$ are implicitly assumed to be independent and to share a common rate parameter. Under these assumptions, the ratio $b = M / (M + U)$ can be shown to follow a beta distribution (\cite{devroye}, Chapter 9). The independence assumption, however, cannot be confirmed by empirical findings, which show that the signal intensities obtained from the Illumina 450k array are often positively correlated (see \cite{Laird}).
	
	\subsection{A Statistical Model for the Ratio of Correlated Gamma Distributed Random Variables}
	\label{s23}
	
	To overcome the problems associated with M-value regression and beta regression, we propose a statistical model (``Ratio of Correlated Gammas (RCG) model'') that is based on the bivariate distribution of the signal intensities $M$ and $U$. In contrast to beta regression, we assume that $M$ and~$U$ are not independent but can be described by a bivariate gamma distribution with probability density function
	\begin{eqnarray}
	\label{kibble}
	\hspace{-.3cm}f_{M,U}(m,u)&\hspace{-.1cm}=\hspace{-.1cm}& \frac{(\lambda_m\lambda_u)^\alpha}{(1-\rho)\, \Gamma(\alpha)}\left(\frac{mu}{\rho\,\lambda_m\lambda_u}\right)^{\frac{\alpha-1}{2}} \exp{\left(-\frac{\lambda_m m}{1-\rho}\right)} \, \nonumber\\[.1cm]
	&\hspace{-.1cm}&\times\,  \exp{\left(-\frac{\lambda_u u}{1-\rho}\right)} I_{\alpha-1}\left(\frac{2\sqrt{\rho\lambda_m\lambda_u mu}}{1-\rho}\right) \, ,
	\end{eqnarray}
	where $\lambda_m, \lambda_u, \alpha >0$, $0 <\rho<1$, and $I_{\alpha-1}$ is the modified Bessel function of the first kind of order~$\alpha-1$. The distribution in (\ref{kibble}) is due to Kibble \cite{kibble} and is often referred to as ``Wicksell-Kibble bivariate gamma distribution'' \cite{contbivar}. As stated in various articles and monographs (e.g.\@ \cite{mardia}), the marginal densities $f_M$,~$f_U$ of $M$ and~$U$, respectively, are given by
	\begin{eqnarray}
	\label{marginaldensities}
	f_M (m) &=& \frac{\lambda_m}{\Gamma (\alpha )} \left( \lambda_m m \right)^{\alpha - 1} \exp (-\lambda_m m) \, ,  \\ \label{marginaldensities2}
	f_U (u) &=& \frac{\lambda_u}{\Gamma (\alpha )} \left( \lambda_u u \right)^{\alpha - 1} \exp (-\lambda_u u) \, . 
	\end{eqnarray}
	The equations in (\ref{marginaldensities}) and (\ref{marginaldensities2}) imply that $M$ and $U$ are gamma distributed random variables with a common shape parameter $\alpha$ and with means and variances given by $\alpha / \lambda_m$, $\alpha / \lambda_u$ and $\alpha / \lambda_m^2$, $\alpha / \lambda_u^2$, respectively. The restriction to a common shape parameter ensures that all measured signal intensities refer to probability density functions sharing the same basic form. On the other hand, the unequal rate parameters $\lambda_m$ and $\lambda_u$ guarantee sufficient flexibility in modeling the differences in the marginal densities of $M$ and $U$ (see (\ref{prresp}) and (\ref{prresp2})). It can further be shown that the Pearson correlation of $M$ and $U$ is equal to $\rho$, implying that (\ref{kibble}) imposes a correlation structure on the two signal intensities (see \cite{contbivar}).
	
	In the next step, the distribution of the ratio $b = M / (M + U)$ is derived:\\
	
	\noindent  {\bf Proposition 1.} Let the distribution of $(M, U)$ be defined by the probability density function in~(\ref{kibble}). Then the ratio $b = M / (M + U)$ follows a univariate distribution with probability density function
	\begin{eqnarray}
	\label{ratiodensity}
	\hspace{-.4cm}f_b (b) &\hspace{-.1cm} =\hspace{-.1cm} & \frac{\Gamma(2\alpha)}{\Gamma^2(\alpha)} \, (\lambda_m\lambda_u)^\alpha \, (1-\rho)^\alpha \, \left(b(1-b)\right)^{\alpha-1} \nonumber \\ 
	&\hspace{-.1cm}& \times\, \frac{\left(\lambda_m b+\lambda_u (1-b)\right)}{\left(\left(\lambda_m b+\lambda_u (1-b)\right)^2-4\rho\lambda_m\lambda_u b(1-b)\right)^{\alpha+0.5}} \ .
	\end{eqnarray}
	
	\noindent {\em Proof:} The proof of Proposition 1, which is related to the work of Nadarajah and Kotz \cite{jscs}, is given in the appendix.\\
	
	The result stated in Proposition 1 can be used to derive the log-likelihood function of a sample of beta values $b_1, \ldots , b_n$:\\
	
	\noindent  {\bf Proposition 2.} For independent sample values $b_1, \ldots , b_n$, the log-likelihood function derived from~(\ref{ratiodensity}) is given by
	\begin{eqnarray}
	\sum_{i=1}^{n} \log(f_b (b_i; \alpha, \rho, \theta))  &=&  \sum_{i=1}^{n} \Big[ \log(\Gamma(2\alpha))
	-2\log (\Gamma(\alpha))	 + \, \alpha\log(\theta) +\,\alpha\log(1-\rho) \nonumber\\
	&& \,  +\log\left((\theta-1)b_i + 1\right) + \, (\alpha-1)\log(b_i(1-b_i))  \nonumber \\[.1cm]
	&&\, -\, \, \big(\alpha+0.5\big) \log \big(   \left((\theta-1)b_i + 1\right)^2-4\, \rho\,\theta \,b_i (1-b_i) \big) \Big] \label{LL} \, ,
	\end{eqnarray}
	where $\theta:=\lambda_m / \lambda_u$.\\
	
	\noindent {\em Proof:} See appendix.\\
	
	Proposition 2 implies that the log-likelihood function derived from (\ref{ratiodensity}) is a function of the mean ratio $\theta = \lambda_m / \lambda_u = \text{E} (U) / \text{E}(M) $.
	
	To quantify the associations between the covariates $\boldsymbol{X}$ and the signal intensities $M$ and $U$, we consider linear predictors $\boldsymbol{X}^\top\zeta_m$ and $\boldsymbol{X}^\top\zeta_u$, $\zeta_m , \zeta_u \in \mathbb{R}^{p+1}$, that relate the vector $\boldsymbol{X} = (1, X_1, \ldots , X_p)^\top$ to the marginal means $\alpha / \lambda_m$ and $\alpha / \lambda_u$, respectively. A convenient link function that guarantees the positivity of $\lambda_m$ and $\lambda_u$ is the logarithmic transformation, resulting in the predictor-response relationships
	\begin{eqnarray}
	\label{prresp}
	\log (\text{E}(M| \boldsymbol{X})) &=& \log ({\alpha}) - \boldsymbol{X}^\top\zeta_m \, ,\\
	\label{prresp2}
	\log (\text{E}(U| \boldsymbol{X}) ) &=& \log (\alpha ) - \boldsymbol{X}^\top\zeta_u \, ,
	\end{eqnarray}
	with $\log (\lambda_m) = \boldsymbol{X}^\top\zeta_m$ and $\log (\lambda_u ) = \boldsymbol{X}^\top\zeta_u$. Note that the term $\log (\alpha )$ can be incorporated into the intercept terms of the coefficient vectors $\zeta_m = (\zeta_{0m}, \zeta_{1m}, \ldots , \zeta_{pm})^\top$ and $\zeta_u = (\zeta_{0u}, \zeta_{1u}, \ldots , \zeta_{pu})^\top$. The model equations in (\ref{prresp}) and~(\ref{prresp2}) are therefore in line with traditional univariate gamma regression approaches that relate the log-transformed mean of the response variable to a linear function of the predictors.
	
	Defining $\gamma = (\gamma_0, \gamma_1 , \ldots , \gamma_p)^\top := \zeta_m - \zeta_u$, the mean ratio $\text{E} (U|\boldsymbol{X}) / \text{E}(M|\boldsymbol{X})$ can be written as~$\theta |\boldsymbol{X} = \exp (\boldsymbol{X}^\top \gamma)$, and the log-likelihood function of a sample $(b_1, \boldsymbol{X}_1^\top ), \ldots , (b_n, \boldsymbol{X}_n^\top )$ becomes
	\begin{eqnarray}
	\label{LLboost} \hspace{-.7cm}
	\sum_{i=1}^{n} \log(f_b (b_i, \boldsymbol{X}_i; \alpha, \rho, \gamma)) 
	&=& \sum_{i=1}^{n} \Bigg[ \log(\Gamma(2\alpha))-2\log (\Gamma(\alpha))  +\, \alpha\,\Xg \nonumber\\
	&& \hspace{-4.5cm} + \, \alpha\log(1-\rho) + \ \log\left(( \exp ( \Xg ) -1) \, b_i + 1\right)  +  \, (\alpha-1)\, \log(b_i(1-b_i)) \nonumber\\[.1cm]
	&& \hspace{-4.5cm} - \, \left(\alpha+ 0.5 \right) \log \Big(   \left(( \exp ( \Xg )-1) \, b_i + 1\right)^2 
	-\, 4 \, \rho \exp (\Xg )\, b_i(1-b_i)   \Big) \Bigg] \, . 
	\end{eqnarray}
	Equations (\ref{prresp}) to (\ref{LLboost}) define a statistical model in which the association between the methylation status $b$ and the covariates $\boldsymbol{X}$ is quantified by the coefficient vector~$\gamma$. If $\gamma_k= 0$, $k \in \{1, \ldots , p\}$, the predictor-response relationships in (\ref{prresp}) and (\ref{prresp2}) imply that $\zeta_{km} = \zeta_{ku}$ and  $\text{E} (M|\boldsymbol{X}) = \text{E} (U|\boldsymbol{X})$ (provided that the values of the other covariates remain constant). Hence, if $\gamma_k= 0$, the $k$-th covariate~$X_k$ has the same effect on both $M$ and $U$, implying that $X_k$ is not associated with the methylation status at the CpG site under consideration. On the other hand, large values of $|\gamma_k |$ result from large differences in the coefficients $\zeta_{km}$ and $\zeta_{ku}$, implying that DNA methylation varies greatly with the value of $X_k$. Assessing the hypotheses ``$H_0: \gamma_k = 0$ vs. $H_1: \gamma_k \ne 0$'' is therefore equivalent to a statistical test on the association between $b $ and~$X_k$.

	\subsection{Estimation and Hypothesis Tests}
	\label{s24}
	
	To obtain a consistent estimator of the coefficient vector $\gamma$, the log-likelihood function in (\ref{LLboost}) needs to be maximized over both $\gamma$ and the hyperparameters $\alpha$ and $\rho$.
	To this purpose, we propose the application of a gradient boosting algorithm with linear base-learning functions, as described in \cite{bh}. For given data $(b_i, \boldsymbol{X}^\top_i)_{i=1,\ldots , n}$, gradient boosting is a generic optimizer that minimizes a risk function $\mathcal{R} (f , (b_i , \boldsymbol{X}_i^\top)_{i=1,\ldots , n})$ over an unknown prediction function $f(\boldsymbol{X})$, with the only requirement being the existence of the derivative $\partial \mathcal{R} / \partial f$ \cite{hothornMiM}. 
	%Hence the algorithm does not rely on the calculation of the Fisher information matrix, the derivation/estimation of which is mathematically and computationally infeasible for the log-likelihood function (\ref{LLboost}) (\cite{jscs}).
	Because the base-learning functions are chosen to be linear in $\boldsymbol{X}$, the space of the prediction function~$f$ is restricted to the subspace defined by $f(\boldsymbol{X}) = \boldsymbol{X}^\top \gamma$, implying that estimation of $f$ reduces to the estimation of the coefficient vector $\gamma$ (see \cite{mboost} for a detailed description of the algorithm). Furthermore, gradient boosting allows for the additional estimation of the hyperparameters $\alpha$ and $\rho$ \cite{schmid2010}. Maximum likelihood (ML) estimates of $\gamma$, $\alpha$ and $\rho$ can therefore be obtained by setting $\mathcal{R}$ equal to the negative of the log-likelihood in (\ref{LLboost}) and by running gradient boosting until convergence.
	
	By standard maximum likelihood arguments, the hypotheses ``$H_0: \gamma_k = 0$ vs.\@ $H_1: \gamma_k \ne 0$'' can be investigated by plugging the ML estimates $\hat{\gamma}$, $\hat{\alpha}$ and $\hat{\rho}$ in the observed information matrix $J (\alpha , \rho, \gamma) = - \sum_{i=1}^n \partial^2 \log ( f_b (b_i, \boldsymbol{X}_i; \alpha , \rho, \gamma)) / \partial^2 \gamma$ and by calculating the test statistic
	\begin{equation} \label{teststat}
	Z_k = \hat{\gamma_k}\, \big/ \sqrt{J^{-1}_{kk}  (\hat{\alpha}, \hat{\rho}, \hat{\gamma})} \, , \ \ \, k \in \{ 1, \ldots , p\} \, ,
	\end{equation}
	where $J^{-1}_{kk}$ denotes the $k$-th diagonal element of $J^{-1}$. Under the null hypothesis, $Z_k$ is asymptotically standard normally distributed as $n\to\infty$. Details on the calculation of $J$ are given in the appendix.

\section{Discussion}
\label{s4}

The development of statistical models to analyze DNA methylation is the subject of intense and ongoing research \cite{park1, zheng, dolz, bmc3}. In this paper, we proposed a likelihood-based approach to analyze and infer the associations between covariates and methylation levels in Illumina 450k data. In contrast to beta regression, the proposed RCG model accounts for possible correlations between methylated and unmethylated signal intensities, thereby increasing the flexibility of the model in describing the distribution of methylation levels at individual CpG sites.

%The proposed RCG methodology is a likelihood-based approach to analyze and infer the effects of covariates in regression models with bounded response variables. In the context of DNA methylation studies, the RCG model provides an extension of traditional analysis techniques for Illumina 450k data that accounts for possible correlations between methylated and unmethylated signal intensities. In contrast to beta regression, the RCG model is based on a bivariate gamma distribution with covariate-specific rate parameters, increasing the flexibility of the model in describing the distribution of methylation levels at individual CpG sites.

The use of a gradient boosting algorithm to optimize the parameters of the RCG model lays the ground for a variety of additional modeling options. For example, it is straightforward to account for nonlinear covariate effects and to extend the linear predictor in (\ref{LLboost}) by a set of spline functions.
%In higher-dimensional settings, variable selection and shrinkage can be incorporated by controlling the stopping iteration of the boosting algorithm.
Furthermore, it is possible to embed the RCG model in the GAMLSS framework \cite{gamlss2005} and to increase its flexibility by relating the parameters $\alpha$ and $\rho$ to separate linear or additive predictors. For details, see \cite{gamboostLSSMayr} and \cite{gamboostLSSHofner}.

\bibliographystyle{plainnat} % Style BST file (bmc-mathphys, vancouver, spbasic).
\bibliography{LiteratureMethylierung}

% or include bibliography directly:
% \begin{thebibliography}
% \bibitem{b1}
% \end{thebibliography}

\newpage
\section*{Appendix}\vspace{.2cm}

\subsection*{Proof of Proposition 1}

We start with a lemma on the properties of the modified Bessel function of the first kind of order $\nu := \alpha-1$.\newline

\noindent {\bf Lemma 1:} \label{Lemma1}
%% s. http://www.lepp.cornell.edu/~ib38/tmp/reading/Table_of_Integrals_Series_and_Products_Tablicy_Integralov_Summ_Rjadov_I_Proizvedennij_Engl._2.pdf, %S. 699, 6.621  ODER
%% https://books.google.de/books?id=2t2cNs00aTgC&printsec=frontcover&hl=de&source=gbs_ge_summary_r&cad=0#v=onepage&q&f=false S.303   (2.15.3.2)
For $\tilde{\alpha}+\nu>0$ and $p>c$ it holds that
\begin{align*}
	\int_{0}^{\infty} x^{\tilde{\alpha}-1}\exp(-px)\, I_{\nu}(cx) \,dx \,=\, p^{-(\tilde{\alpha}+\nu)} \left(\frac{c}{2}\right)^\nu \frac{\Gamma(\nu+\tilde{\alpha})}{\Gamma(\nu+1)}
	\text{ }_2F_1\left(\frac{\nu+\tilde{\alpha}}{2}, \frac{\nu+\tilde{\alpha}+1}{2}, \nu+1, \frac{c^2}{p^2}\right) \, ,
\end{align*}
where $\text{}_2F_1 (\cdot )$ is the Gauss hypergeometric function (see \cite{jscs}, p.\@ 350). For a formal proof of Lemma 1, see~\cite{prudnikov}. \qed\newline

\noindent The proof of Proposition 1 is obtained by deriving the joint density function $f_{R,b}$ of the random variables $R :=M+U$ and $b=M / (M+U) =M /R $. Transforming $(M,U) = (Rb, R(1-b))$ into~$(R,b)$ yields the Jacobian matrix
\begin{equation}
	\tilde{J} = \left( \begin{array}{cc}
		\frac{\partial Rb}{\partial R} & \frac{\partial Rb}{\partial b} \\[.1cm]
		\frac{\partial R (1-b)}{\partial R} & \frac{\partial R  (1-b)}{\partial b}
	\end{array}
	\right)
	= \left( \begin{array}{cc}
		b & R \\
		(1-b) & -R
	\end{array}
	\right)
\end{equation}
with $|\text{det}(\tilde{J})| = R$. It follows that, under the assumptions of Proposition 1,
%$f_{R,b}$ is given by
\begin{eqnarray}
	f_{R,b}(r,b)%&=r\frac{(\lambda_M\lambda_U)^\alpha}{(1-\rho)\Gamma(\alpha)}\left(\frac{(r^2w(1-w))}{\rho\lambda_M\lambda_U}\right)^\frac{\alpha-1}{2}
	%\exp\left(-\frac{\lambda_Mrw+\lambda_Ur(1-w)}{1-\rho}\right)I_{\alpha-1}\left(\frac{2*\sqrt{\rho\lambda_M\lambda_Ur^2w(1-w)}}{1-\rho}\right)\\
	&=&\frac{(\lambda_m\lambda_u)^{\frac{\alpha+1}{2}}}{(1-\rho)\,\rho^{\frac{\alpha-1}{2}}\Gamma(\alpha)} \ r^\alpha \left(b \, (1-b)\right)^\frac{\alpha-1}{2}
	\exp\left(-\frac{\lambda_m rb+\lambda_u r(1-b)}{1-\rho}\right) \nonumber
	\\ && \times \, I_{\alpha-1}\left(\frac{2\sqrt{\rho\,\lambda_m \lambda_u \, r^2 \, b \,(1-b)}}{1-\rho}\right)
	\, .
\end{eqnarray}
Defining
\begin{equation}
	Z(b) := \int r^\alpha \exp\left(-\frac{\lambda_m rb+\lambda_u r(1-b)}{1-\rho}\right)  I_{\alpha-1}\left(\frac{2\sqrt{\rho \,\lambda_m \lambda_u \, r^2 \, b \, (1-b)}}{1-\rho}\right)dr \, ,
\end{equation}
the marginal density function $f_b(b)$ is derived by integrating $f_{R,b}$ over $R$:
\begin{align}\label{Wdens1}
	&f_b(b)=\int f_{R,b}(r,b)\, dr = \frac{(\lambda_m\lambda_u )^{\frac{\alpha+1}{2}}}{(1-\rho)\, \rho^{\frac{\alpha-1}{2}}\, \Gamma(\alpha)} \left(b\, (1-b)\right)^\frac{\alpha-1}{2} Z(b) \, .
\end{align}
Setting 
\begin{equation}
	\tilde{\alpha}=\alpha+1 \, , \ \ \nu=\alpha-1 \,  , \ \ p=\frac{\lambda_m b+\lambda_u (1-b)}{1-\rho} \, , \ \ c=\frac{2\sqrt{\rho\, \lambda_m \lambda_u\,  b\, (1-b)}}{1-\rho}
\end{equation}
and making use of the fact that
\begin{equation}
	\text{ }_2F_1\left( \alpha , \delta, \alpha , x \right) = (1 - x)^{-\delta} \, ,
\end{equation}
one obtains by application of Lemma 1 that
\begin{align} \label{proofProp2Zb}
	\hspace{-0.8cm}Z(b)= \frac{\Gamma(2\alpha)}{\Gamma(\alpha)} \, (1-\rho)^{\alpha+1}\frac{\left(\sqrt{\rho \,\lambda_m\lambda_u \, b \, (1-b)}\right)^{\alpha-1}}{(\lambda_m b+\lambda_u (1-b))^{2\alpha}}\left(1-\frac{4\rho \, \lambda_m\lambda_u \, b \, (1-b)}{(\lambda_m b+\lambda_u (1-b))^2}\right)^{-\frac{2\alpha+1}{2}}\, .
\end{align}
Combining (\ref{Wdens1}) and (\ref{proofProp2Zb}) yields the probability density function stated in Proposition 1. \qed
%
%Finally, combining (\ref{Wdens1}) and (\ref{JW}) leads to the density of ratio $W$
%\begin{align}
%f_W(w)=\frac{\Gamma(2\alpha)}{\Gamma^2(\alpha)}(\lambda_M\lambda_U)^\alpha (1-\rho)^\alpha\frac{\left(\lambda_M w+\lambda_M (1-w)\right)\left(w(1-w)\right)^{\alpha-1}}{\left(\left(\lambda_M w+\lambda_U (1-w)\right)^2-4\rho\lambda_M\lambda_U w(1-w)\right)^{\alpha+\frac{1}{2}}}
%\end{align}
%for $0<w<1$.
%
%\textsc{}

\subsection*{Proof of Proposition 2}

Defining $\theta := \lambda_m / \lambda_u$, the log-likelihood function derived from of Equation (9) of the manuscript becomes
\begin{eqnarray}
	\sum_{i=1}^{n} \log(f_b (b_i; \alpha, \rho, \theta)) &=& \sum_{i=1}^{n} \Big[ \log (\Gamma (2\alpha)) - 2 \log (\Gamma (\alpha)) + \alpha \log (\lambda_m \lambda_u) + \alpha \log (1-\rho) \nonumber\\
	&& +\,(\alpha - 1) \log(b_i(1-b_i)) + \log (\lambda_m b_i + \lambda_u (1-b_i)) \nonumber\\
	&& - \, (\alpha + 0.5) \log \left( (\lambda_m b_i + \lambda_u (1-b_i))^2 - 4\, \rho \, \lambda_m
	\lambda_u \, b_i (1-b_i)\right) \Big] \nonumber\\
	&=& \sum_{i=1}^{n} \Big[ \log (\Gamma (2\alpha)) - 2 \log (\Gamma (\alpha)) + \alpha \log (\theta \lambda_u^2)+ \alpha \log (1-\rho) \nonumber\\
	&& +\,(\alpha - 1) \log(b_i(1-b_i)) + \log \left( \lambda_u \,(( \theta - 1 ) \, b_i + 1 )\right) \nonumber\\
	&& - \, (\alpha + 0.5) \log \left( ( \lambda_u ((\theta - 1) \, b_i + 1))^2 - 4\, \rho \, \theta
	\lambda_u^2 \, b_i (1-b_i)\right) \Big] \nonumber\\
	&=& \sum_{i=1}^{n} \Big[ \log (\Gamma (2\alpha)) - 2 \log (\Gamma (\alpha)) + \alpha \log (\theta )+ \alpha \log (1-\rho) \nonumber\\
	&& +\,(\alpha - 1) \log(b_i(1-b_i)) + \log \left( ( \theta - 1 ) \, b_i + 1 \right) \nonumber\\
	&& - \, (\alpha + 0.5) \log \left( ( (\theta - 1) \, b_i + 1)^2 - 4\, \rho \, \theta \, b_i (1-b_i)\right) \Big] \, .
\end{eqnarray} \qed

\subsection*{Derivation of the Observed Information Matrix}

Defining $D_1 := \denomS$ and $D_2 := \denomL$, the first derivative of Equation~(13) of the manuscript w.r.t.\@ $\gamma$ is given by
\begin{eqnarray}
&&\frac{\partial}{\partial \gamma} \, \sum_{i=1}^{n} \log(f_b(b_i, \boldsymbol{X}_i ; \alpha, \rho , \gamma)) \nonumber \\&&=
 \, \sum_{i=1}^{n}  \Big[ \alpha\X + \frac{\X b_i\eXg}{D_1}-\left(\alpha+0.5\right) 
\frac{ 2 \left( D_1 -2 \, \rho \, (1-b_i)\right) \X b_i \eXg }{D_2} \,\Big] \, .
%= \sum_{i=1}^{n} & \alpha\X + \frac{\X w_i\eXg}{D_1}-\left(\alpha+\frac{1}{2}\right) 
%\frac{2D_1 \X w_i \eXg - 4\rho \X \eXg w_i (1-w_i)}{D_2}
\end{eqnarray}
It follows that the observed information matrix is given by
\begin{eqnarray}
\hspace{-.5cm} J (\alpha, \rho , \gamma) &=& -\frac{\partial^2}{\partial^2 \gamma} \, \sum_{i=1}^{n} \log(f_b(b_i, \boldsymbol{X}_i ; \alpha, \rho , \gamma)) \\
&&= \ \sum_{i=1}^{n} \Big[ \,\frac{ \left( D_1 - b_i \eXg \right) b_i \eXg \XtX}{D_1^2}\nonumber\\ 
&& \ \ \ - \left(\alpha+0.5 \right) \, \frac{2 \, b_i \eXg \left( D_1 + b_i \eXg
 	- 2 \,\rho\, (1-b_i) \right) \XtX}{D_2}	\nonumber\\
 &&\ \ \ + \left(\alpha+0.5\right) \, \frac{4 \, b_i^2 \exp(2\boldsymbol{X}_i^\top \gamma) \left( D_1-2 \rho (1-b_i) \right)^2 \XtX}{D_2^2} \, \Big] \, .
\end{eqnarray}
%= \sum_{i=1}^{n} &  \frac{D_1 \XtX w_i \eXg -\left( w_i\X \eXg \right)^2}{D_1^2} \\
%& -  \frac{\left(\alpha+\frac{1}{2}\right)}{D_2^2} 
%\Bigg[2w_i\eXg \XtX \left(w_i\eXg+D_1-2\rho(1-w_i)\right)D_2\Bigg] \\
%& +  \frac{\left(\alpha+\frac{1}{2}\right)}{D_2^2} \Bigg[ - \left(2\X w_i \eXg \left( D_1-2\rho(1-w_i) \right)\right)^2\Bigg] \\

\end{document}